\documentclass[12pt]{article}

\usepackage{amsmath,amssymb}
\usepackage{graphicx}

\makeatletter
 
  \@addtoreset{equation}{section}
 \makeatother
\newcommand{\bfa}{{\mathbf a}}

\newcommand{\bfe}{{\mathbf e}}

\newcommand{\bfn}{{\mathbf n}}

\newcommand{\ba}{\bar{a}}
\newcommand{\bb}{\bar{b}}
\newcommand{\bz}{\bar{z}}

\newcommand{\cD}{{\cal D}}

\newcommand{\cN}{{\cal N}}
\newcommand{\cO}{{\cal O}}

\newcommand{\Z}{\mathbb{Z}}
\newcommand{\R}{\mathbb{R}}
\newcommand{\C}{\mathbb{C}}

\newcommand{\nn}{\nonumber}

\newcommand{\Tr}{{\rm Tr}\,}

\newcommand{\e}{\epsilon}

\begin{document}

\begin{titlepage}

\setcounter{page}{0}
\renewcommand{\thefootnote}{\fnsymbol{footnote}}

\vspace{20mm}

\begin{center}
{\large\bf
Exact Vacuum Energy of Orbifold Lattice Theories}

\vspace{20mm}
{
So Matsuura\footnote{{\tt matsuura@nbi.dk}}}\\
\vspace{10mm}

{\em The Niels Bohr Institute, \\
The Niels Bohr International Academy,\\
Blegdamsvej 17, DK-2100 Copenhagen, Denmark}

\end{center}

\vspace{20mm}
\centerline{{\bf Abstract}}
\vspace{10mm}

We investigate the orbifold lattice theories constructed from 
supersymmetric Yang-Mills matrix theories (mother theories) 
with four and eight supercharges. We show that 
the vacuum energy of these theories 
does not receive any quantum correction perturbatively. 

\end{titlepage}
\newpage

\renewcommand{\thefootnote}{\arabic{footnote}}
\setcounter{footnote}{0}

\section{Introduction}

Recently, there has been a rapid development in 
supersymmetric lattice gauge theories. 
A systematic way to construct supersymmetric lattice formulations 
is developed in 
\cite{Kaplan:2002wv}--\nocite{Cohen:2003xe}%
\nocite{Cohen:2003qw}\cite{Kaplan:2005ta}, 
where 
a space-time lattice is generated by an orbifold projection 
of a supersymmetric Yang-Mills matrix theory (mother theory), 
and a lattice spacing is introduced by ``deconstruction'' 
\cite{Arkani-Hamed:2001ca}. 
By choosing the orbifold projection properly, one can make 
at least one supercharge or BRST charge preserved 
on the lattice. 
These formulations are further analysed in 
\cite{Giedt:2003xr}--\nocite{Giedt:2003ve,Onogi:2005cz}%
\nocite{Ohta:2006qz}\cite{Damgaard:2007be}%
\footnote{
For a nice review, see \cite{Giedt:2006pd}.}. 
A prescription to generate a lattice theory from a topologically twisted 
continuum supersymmetric gauge theory is proposed by Catterall  
\cite{Catterall:2003wd}--\nocite{Catterall:2004np}\cite{Catterall:2005fd}.
In these formulations, 
the BRST charge of the continuum theory 
is preserved on the lattice. 
A characteristic feature of these formulations is that 
all the degrees of freedom on the lattice
except for site variables are doubled by a complexification 
and the path-integral is performed along ``the real line''. 
Numerical simulations are carried out 
for the model of two-dimensional $\cN=(2,2)$ supersymmetric gauge
theory \cite{Catterall:2006jw}, 
which reproduce the Ward-Takahashi identities in fairly good accuracy. 
Other formulations constructed from topologically twisted 
supersymmetric gauge theories are developed by Sugino
\cite{Sugino:2003yb}--\nocite{Sugino:2004qd}%
\nocite{Sugino:2004uv}\cite{Sugino:2006uf}, 
where it is shown that the BRST transformation for the continuum fields 
can also be defined for lattice variables. 
The lattice action is straightforwardly generated from the $Q$-exact 
form of the continuum action by replacing all the fields by the lattice 
variables. 
A common feature of the above three 
formulations is that they possess at least 
one preserved supercharge or BRST charge. 
Alternative approach (the link approach) has been developed in 
\cite{D'Adda:2004jb}--\nocite{D'Adda:2005zk}\cite{D'Adda:2007ax}, 
where it is claimed that all the supersymmetry of the 
continuum theory is preserved 
on the lattice. 
They first explicitly construct a supersymmetry algebra 
on a lattice and next make a lattice action based on the algebra, 
although there are some discussions on this approach 
\cite{Bruckmann:2006ub}\cite{Bruckmann:2006kb}. 
For conventional but useful approaches to supersymmetric 
lattice gauge theories, see 
\cite{Nishimura:1997vg}--\nocite{Kaplan:1983sk}\nocite{Maru:1997kh}%
\nocite{Neuberger:1997bg}\nocite{Kaplan:1999jn}\nocite{Fleming:2000fa}%
\cite{Montvay:2001aj} 
in which the theories do not have any supersymmetry on 
a lattice but 
they flow to supersymmetric theories without fine-tuning 
because of a discrete chiral symmetry on the lattice. 
See also \cite{Suzuki:2007jt} 
for a recent lattice approach to two-dimensional 
$\cN=(2,2)$ supersymmetric gauge theory.

The above seemingly different supersymmetric lattice formulations 
with a supercharge on the lattice  
are related to the orbifold lattice theories.  
In fact, the prescription given by Catterall can be  
reproduced using the orbifolding procedure \cite{Damgaard:2007xi}. 
Sugino's formulations can be obtained 
from Catterall's formulations by restricting the degrees of freedom of 
the complexified fields with preserving the supercharge 
\cite{Takimi:2007nn}. 
Furthermore, the formulations given by the link approach have been
shown to be equivalent to those given by orbifolding
\cite{Damgaard:2007eh}. 
In this sense, it seems important to examine quantum mechanical 
properties of the orbifold lattice theories. 
In the next section, we examine the vacuum energy of 
the orbifold lattice theories constructed from $Q=4$ 
and $Q=8$ mother theories. 
We show that the vacuum energy exactly vanishes to all orders 
of the perturbation theory and 
the flat directions of these theories are 
never lifted up by any perturbative effect. 
The final section is devoted to conclusion and discussion.

\section{Quantum Corrections to Vacuum Energy}

\subsection{Orbifold lattice theories from $Q=4$ mother theory}
\label{Q=4}

As discussed in detail in 
\cite{Kaplan:2002wv}--\nocite{Cohen:2003xe}\nocite{Cohen:2003qw}%
\cite{Kaplan:2005ta}, 
an orbifold lattice theory is obtained 
by performing an appropriate orbifold projection to 
a supersymmetric Yang-Mills matrix theory (mother theory) followed by 
deconstruction, that is, by expanding the orbifolded matrix theory
around a classical vacuum. 
Let us start with the orbifold lattice theories constructed from 
the dimensionally reduced four-dimensional $\cN=1$ 
supersymmetric Yang-Mills theory \cite{Cohen:2003xe}. 
As discussed in \cite{Damgaard:2007be}, 
the lattice gauge theory obtained from this mother theory 
is essentially unique to be a lattice formulation for 
two-dimensional $\cN=(2,2)$ supersymmetric Yang-Mills theory%
\footnote{
In this paper, we restrict ourselves to consider gauge theories 
in $d$ dimensional space-time with $d\ge 2$. 
}.
The action of the orbifolded matrix theory (before deconstruction) 
is given by
\begin{align}
S_{\rm orb} &=
\frac{1}{g^2}{\rm Tr}\sum_{\bfn \in \Z_N^2}\Biggl( 
\frac{1}{4}\Bigl|z_m({\bfn})
z_n(\bfn+\bfe_m) - z_n(\bfn)z_m(\bfn+\bfe_n)\Bigr|^2  \cr
&
 + \frac{1}{8}\Bigl(z_m(\bfn)\bar{z}_m(\bfn)-\bar{z}_m(\bfn-\bfe_m)z_m(\bfn
-\bfe_m)\Bigr)^2 \cr
& + \psi_m(\bfn)\Bigl(\bar{z}_m(\bfn)\eta(\bfn)-\eta(\bfn+\bfe_m)
\bar{z}_m(\bfn)\Bigr)\cr 
& - \frac{1}{2}\chi_{mn}(\bfn)\Bigl(z_m(\bfn)\psi_n(\bfn+\bfe_n)
-\psi_n(\bfn)z_m(\bfn+\bfe_n)
-(m\leftrightarrow n)\Bigr)\Biggr) ~, 
\label{orbifold action Q=4}
\end{align}
where $m,n=1,2$, 
$\bfe_m$ are two linearly independent 
integer valued two-vectors, 
and all the fields are complex matrices with the size $M$. 
Although this action does not contain any lattice spacing nor kinetic
terms, we can regard it as a lattice action
by identifying $\bfn$ as the label of a site on a two-dimensional 
square lattice with the size $N$. 
In this sense, 
the variables $z_m(\bfn)$ and $\bz_m(\bfn)$ are bosonic fields 
living on the links $(\bfn,\bfn+\bfe_m)$ and $(\bfn+\bfe_m,\bfn)$, 
respectively, and $\eta(\bfn)$, $\psi_m(\bfn)$ and 
$\chi_{12}(\bfn)=-\chi_{21}(\bfn)$ 
are fermionic fields living on the site $\bfn$, 
the link $(\bfn,\bfn+\bfe_m)$ and the link 
$(\bfn+\bfe_1+\bfe_2,\bfn)$, respectively. 
Note the action (\ref{orbifold action Q=4}) is invariant under 
a $U(M)$ ``gauge transformation'' 
$z_m(\bfn)\to g^{-1}(\bfn)z_m(\bfn)g(\bfn+\bfe_m)$ $(g(\bfn)\in U(M))$, 
and so on. 
As mentioned above, kinetic terms and a lattice spacing $a$ are 
introduced by expanding $z_m(\bfn)$ and $\bz_m(\bfn)$ as 
\begin{equation}
 z_m(\bfn) = \frac{1}{a}{\mathbf 1}_M + z'_m(\bfn), 
\quad \bz_m(\bfn) = \frac{1}{a}{\mathbf 1}_M + \bz'_m(\bfn), 
\label{shift}
\end{equation}
then we obtain a lattice formulation for two-dimensional 
$\cN=(2,2)$ supersymmetric Yang-Mills theory. 
Since the potential terms of this theory are given by 
\begin{equation}
 \frac{1}{4}\Bigl|z'_m({\bfn})
z'_n(\bfn+\bfe_m) - z'_n(\bfn)z'_m(\bfn+\bfe_n)\Bigr|^2
 + \frac{1}{8}\Bigl(z'_m(\bfn)\bar{z'}_m(\bfn)-\bar{z'}_m(\bfn-\bfe_m)z'_m(\bfn
-\bfe_m)\Bigr)^2, 
\end{equation}
the classical moduli space (the flat directions) 
of this theory is parametrized by 
the vacuum expectation values of $z'_m(\bfn)$ and $\bz'_m(\bfn)$, 
\begin{equation}
 z'_m(\bfn)=\left(
\begin{matrix}b_m^1 & & \\ & \ddots & \\ && b_m^M\end{matrix}
\right)\equiv b_m, \qquad 
\bz'_m(\bfn)=\left(
\begin{matrix}\bar{b}_m^1 & & \\ & \ddots & \\ && \bar{b}_m^M\end{matrix}
\right)\equiv \bar{b}_m, 
\label{lattice vacuum}
\end{equation}
with $b_m^i\in \C$ ($i=1,\cdots,M$) up to gauge transformations.

In this paper, we are interested in quantum corrections 
to this classical moduli space. 
To examine them, we will estimate the vacuum energy 
at the point (\ref{lattice vacuum}) in the classical moduli space. 
Perturbatively, 
this is achieved by expanding the lattice action (after deconstruction) 
around the vacuum 
(\ref{lattice vacuum}) and summing up all 1PI vacuum graphs. 
However, recalling that the action of the lattice gauge theory is 
obtained by substituting (\ref{shift}) into the action of 
the orbifolded matrix theory (\ref{orbifold action Q=4}), 
we see that the same result is obtained by directly 
replacing $z_m(\bfn)$ and $\bz_m(\bfn)$ 
in the action (\ref{orbifold action Q=4}) with 
\begin{align}
 z_m(\bfn)&\to z_m(\bfn)+\frac{1}{a}{\mathbf 1}_M + b_m
\equiv z_m(\bfn)+  a_m, \nn \\ 
 \bz_m(\bfn)&\to \bz_m(\bfn)+\frac{1}{a}{\mathbf 1}_M + \bb_m
\equiv \bz_m(\bfn)+\ba_m, 
\label{comment}
\end{align}
respectively. 
In the following calculation, we will use this notation 
and estimate the vacuum energy as a function of $a_m^i$ $(i=1,\cdots,M)$.

We first calculate the 1-loop vacuum energy. 
It is convenient to fix the gauge by imposing a gauge condition, 
\begin{equation}
 D^-_m \bz_m(\bfn) - \bar{D}^-_m z_m(\bfn) =0, 
\label{gauge fix}
\end{equation}
where 
\begin{align}
D^-_m f(\bfn)\equiv
 a_m f(\bfn) - f(\bfn-\bfe_m) a_m, \qquad
\bar{D}^-_m f(\bfn) \equiv
-\ba_m f(\bfn-\bfe_m) + f(\bfn) \ba_m. 
\label{D-}
\end{align}
For the purpose of the later discussion, we also define 
\begin{align}
D^+_m f(\bfn)\equiv
 a_m f(\bfn+\bfe_m) - f(\bfn) a_m, \qquad
\bar{D}^+_m f(\bfn)\equiv
-\ba_m f(\bfn) + f(\bfn+\bfe_m) \ba_m. 
\label{D+}
\end{align}
By introducing gauge fixing terms and FP ghost fields 
corresponding to the gauge condition (\ref{gauge fix}) 
in a standard way, the second-order action is obtained as
\begin{align}
 S^{(2)} = \frac{1}{g^2}\Tr \sum_{\bfn\in\Z_N^2}\Biggl(
&\frac{1}{2}\bar{D}^-_n z_m(\bfn) D^-_n z_m(\bfn)
+\frac{1}{2}\bar{D}^-_n B(\bfn) D^-_n C(\bfn) \nn \\
&+\eta(\bfn) \bar{D}^-_m \psi_m(\bfn)
-\frac{1}{2}\chi_{mn}(\bfn)\Bigl(D^+_m \psi_n(\bfn)- D^+_n \psi_m(\bfn) \Bigr)
\Biggr), 
\label{2nd order action}
\end{align}
where $B(\bfn)$ and $C(\bfn)$ are FP ghost fields. 
By integrating over the fields, we get the 1-loop contribution 
to the partition function as%
\footnote{
In this calculation, the constant modes 
are treated by shifting the difference operators 
(\ref{D-}) and (\ref{D+}) as $D_m^\pm \to D_m^\pm + i\mu$, which 
corresponds to adding mass terms as done in
\cite{Cohen:2003xe}. 
Although this modification breaks the BRST symmetry, 
the final result of the following discussion still holds 
in the limit of $\mu\to 0$ since the breaking of the 
symmetry is soft. 
} 
\begin{align}
 Z\Bigr|_{\rm 1-loop} &= \int \prod_{\bfn}d\Phi(\bfn) 
e^{-S^{(2)}[\Phi(\bfn)]} \nn \\
&= \frac{\det \Delta}{\det \Delta} =1, 
\label{1-loop result}
\end{align}
where $\Delta\equiv \sum_m \bar{D}_m^+ D_m^-$ is the 
Laplacian and the lattice variables have been abbreviated as 
$\Phi(\bfn)$. 
The denominator of the second line comes from the contributions 
from the bosonic fields and the ghost fields and 
the numerator comes from the fermionic fields. 
The result (\ref{1-loop result}) means that
the vacuum energy is equal to zero and 
the classical flat directions remain flat at the 1-loop level. 
Note that the same calculation is carried out at the origin of 
the moduli space in \cite{Onogi:2005cz}. 
We can reproduce it by setting $b_m^i=0$ (or $a_m^i=1/a$) 
in our calculation.

One might think that, 
even though the 1-loop contribution to the vacuum energy is zero, 
higher-loop contributions 
would give non-trivial corrections to the vacuum energy, 
since the supersymmetry is almost broken except for 
the only one preserved supercharge (or BRST charge). 
However, we can show that it is not the case and 
the above 1-loop result is exact. 
The key point is that the action (\ref{orbifold action Q=4}) can be 
written in a $Q$-exact form \cite{Cohen:2003xe}:
\begin{align}
S_{\rm orb} =
\frac{1}{2g^2}{\rm Tr}\sum_{\bfn \in \Z_N^2}Q\Biggl( 
& \eta(\bfn)\Bigl({z}_m(\bfn) \bz_m(\bfn)
-\bz_m(\bfn-\bfe_m){z}_m(\bfn-\bfe_m)+d(\bfn)\Bigr) \nn \\
& - \chi_{mn}(\bfn)\Bigl(z_m(\bfn)z_n(\bfn+\bfe_n)
-z_n(\bfn)z_m(\bfn+\bfe_n)\Bigr)\Biggr) ~, 
\label{Q-exact action Q=4}
\end{align}
where $Q$ is a BRST charge that acts on the fields as 
\begin{align}
 Q z_m(\bfn) &= \psi_m(\bfn), \quad Q \bz_m(\bfn) = 0, \nn \\
 Q d(\bfn) &= \psi_m(\bfn)\bz_m(\bfn)
   -\bz_m(\bfn-\bfe_m)\psi_m(\bfn-\bfe_m), \nn \\
\label{BRST Q=4}
 Q\eta(\bfn) &=
 \frac{1}{4}\Bigl(z_m(\bfn)\bz_m(\bfn)
  -\bz_m(\bfn-\bfe_m)z_m(\bfn-\bfe_m)-d(\bfn)\Bigr), \\
 Q\chi_{mn}(\bfn) &= \frac{1}{2}\Bigl(\bz_m(\bfn+\bfe_n)\bz_n(\bfn)
  -\bz_n(\bfn+\bfe_m)\bz_m(\bfn)\Bigr), \nn 
\end{align}
and $d(\bfn)$ is an auxiliary bosonic field which 
makes $Q$ be nilpotent off-shell. 
Recalling the discussion in topological field theory
\cite{Witten:1988ze}, we see that the partition function 
of this theory does not depend on the coupling constant $g$. 
In fact, if we write the partition function as 
$Z(g)=\int \cD \Phi e^{\frac{1}{g^2}Q\Xi[\Phi]}$, 
the derivative of the partition function by $g$ gives 
\begin{equation}
 \frac{d}{dg} Z(g) \propto \Bigl\langle Q\Xi[\Phi] \Bigr\rangle=0, 
  \label{g-independence}
\end{equation}
where $\langle \cO \rangle$ denotes the expectation value of 
an operator $\cO$ and we have used 
the fact that, 
as long as the BRST symmetry is not broken spontaneously, 
the expectation value of a $Q$-exact operator vanishes 
when the action is $Q$-exact. 
This means that 
the partition function evaluated in the weak coupling 
limit, that is, the 1-loop result given above is exact. 
In particular, we can expect that all the higher-loop contributions 
to the vacuum energy vanish.

Note that one might think that the partition function given above 
expresses not the vacuum 
energy but the Witten index of the theory since we impose the periodic 
boundary condition to the fermionic fields in the time direction. 
Although it is actually the case, 
the boundary conditions do not affect the perturbative contributions 
in the limit that the period of the time direction goes to infinity. 
Therefore we can conclude that there is no perturbative correction 
to the vacuum energy from (\ref{g-independence}).%
\footnote{The author would like to thank to 
H.~Suzuki for discussing this point. 
}

Another note is that 
we can apply the same analysis to a deformed theory given by%
\footnote{The physical interpretation of this deformation is 
still unclear.
In fact, the continuum limit of this deformed theory is not 
Lorentz invariant, though it has a BRST symmetry generated by $Q$. 
The author would like to thank M.~\"{U}nsal 
for pointing it out. } 
\begin{align}
S_{\rm orb} =
\frac{1}{2g^2}{\rm Tr}\sum_{\bfn \in \Z_N^2}Q\Biggl( 
& \eta(\bfn)\Bigl({z}_m(\bfn) \bz_m(\bfn)
-\bz_m(\bfn-\bfe_m){z}_m(\bfn-\bfe_m)+d(\bfn)\Bigr) \nn \\
& -\beta \chi_{mn}(\bfn)\Bigl(z_m(\bfn)z_n(\bfn+\bfe_n)
-z_n(\bfn)z_m(\bfn+\bfe_n)\Bigr)\Biggr) ~, 
\label{deformed Q-exact action Q=4}
\end{align}
where $\beta \in \R$ and the BRST transformation is given by 
(\ref{BRST Q=4}). 
By construction, this deformation does not spoil the 
$Q$-exactness of the action 
and it becomes identical with  
the original orbifolded matrix theory (\ref{orbifold action Q=4}) 
by setting $\beta=1$. 
We can show that the vacuum energy of this deformed 
theory also vanishes at the 1-loop level. 
Therefore, repeating the same argument above, 
we can conclude that there is no perturbative correction 
to the vacuum energy of this theory.

\subsection{Orbifold lattice theories from $Q=8$ mother theory}
\label{Q=8}

Next we consider the lattice theories constructed from 
the mother theory with eight supercharges, that is, 
the dimensionally reduced six-dimensional $\cN=1$ supersymmetric 
Yang-Mills theory \cite{Cohen:2003qw}. 
By performing an orbifold projection to the mother theory, 
we obtain the action of the orbifolded matrix theory 
\cite{Damgaard:2007be}:
\begin{align}
S_{\rm orb} &= \frac{1}{g^2}{\rm Tr}\sum_{\bfn\in\Z_N^d}\Biggl( 
\frac{1}{4}\Bigl| z_m({\bfn})
z_n(\bfn+\bfe_m) - z_n(\bfn)z_m(\bfn+\bfe_n)\Bigr|^2  \cr
& + \frac{1}{8}\Bigl(z_m(\bfn)\bz_m(\bfn)
-\bz_m(\bfn-\bfe_m)z_m(\bfn
-\bfe_m)\Bigr)^2 \cr
& - \psi_m(\bfn)\Bigl(\bz_m(\bfn)\eta(\bfn)-\eta(\bfn+\bfe_m)
\bz_m(\bfn)\Bigr)\cr 
& +\frac{1}{2} \xi_{mn}(\bfn)\Bigl(z_m(\bfn)\psi_n(\bfn+\bfe_m)
-\psi_n(\bfn)z_m(\bfn+\bfe_n)-(m\!\leftrightarrow\!n)\Bigr) \cr
& -\frac{1}{2}\chi_{lmn}(\bfn)\Bigl(
\bz_l(\bfn+\bfe_m+\bfe_n)\xi_{mn}(\bfn)
-\xi_{mn}(\bfn+\bfe_l)\bz_l(\bfn)
\Bigr)\Biggr)~,  
\label{orbifold action Q=8}
\end{align}
where $l,m,n=1,2,3$, $\bfe_m$ are integer valued three-component 
vectors, $d$ is the number of linearly independent vectors
in $\{\bfe_m\}$, and again we assume that all the fields are 
complex matrices with the size $M$. 
Note that $d$ is the maximal dimensionality of the 
lattice theory obtained after deconstruction. 
The fields $z_m(\bfn)$ and $\bz_m(\bfn)$
are bosonic fields living on links $(\bfn,\bfn+\bfe_m)$ and
$(\bfn+\bfe_m,\bfn)$, respectively, and $\eta(\bfn)$, $\psi_m(\bfn)$, 
$\xi_{mn}(\bfn)$ and
$\chi_{lmn}(\bfn)$ are fermionic fields on the site $\bfn$, the link
$(\bfn,\bfn+\bfe_m)$, the link 
$(\bfn+\bfe_m+\bfe_n,\bfn)$ and the link
$(\bfn,\bfn+\bfe_l+\bfe_m+\bfe_n)$, respectively. 
The fields $\xi_{mn}(\bfn)$ and $\chi_{lmn}(\bfn)$ are 
antisymmetric in terms of a permutation of the indices.

In this case, 
we can construct several kinds of supersymmetric 
lattice gauge theories with a different dimensionality, 
with a different number of preserved supercharges 
and with a different lattice structure 
by changing the vectors $\bfe_m$ and the number of bosonic fields 
to shift as (\ref{shift}) \cite{Damgaard:2007be}. 
Recalling the discussion around (\ref{comment}), however, we can estimate 
the vacuum energy of these theories at once 
by directly expanding the orbifolded matrix theory 
(\ref{orbifold action Q=8}) around 
\begin{equation}
 z_m(\bfn)=\left(
\begin{matrix}a_m^1 & & \\ & \ddots & \\ && a_m^M\end{matrix}
\right)\equiv a_m, \qquad 
\bz_m(\bfn)=\left(
\begin{matrix}\bar{a}_m^1 & & \\ & \ddots & \\ && \bar{a}_m^M\end{matrix}
\right)\equiv \bar{a}_m. 
\label{vacuum Q=8}
\end{equation}
By fixing the gauge by the gauge condition (\ref{gauge fix}), 
we obtain the second-order action, 
\begin{align}
 S^{(2)} = \frac{1}{g^2}\Tr \sum_{\bfn\in\Z_N^d}\Biggl(
&\frac{1}{2}\bar{D}^-_n z_m(\bfn) D^-_n z_m(\bfn)
+\frac{1}{2}\bar{D}^-_n B(\bfn) D^-_n C(\bfn) 
+\eta(\bfn) \bar{D}^-_m \psi_m(\bfn)\nn \\ 
&-\frac{1}{2}\xi_{mn}(\bfn)\Bigl(D^+_m \psi_n(\bfn)- D^+_n \psi_m(\bfn) 
\Bigr) 
+\frac{1}{2}\xi_{mn}(\bfn)\bar{D}_l^- \chi_{lmn}(\bfn)
\Biggr). 
\label{2nd order action Q=8}
\end{align}
From this expression, 
it is easy to show that the 1-loop contribution 
to the vacuum energy vanishes again. 

As for the case of the $Q=4$ orbifold lattice theories, 
the lattice theory 
(\ref{orbifold action Q=8}) possesses a BRST charge 
$Q$ that acts on the fields as \cite{Cohen:2003qw}
\begin{align}
 Q z_m(\bfn) &= \psi_m(\bfn), \quad Q \bz_m(\bfn) = 0, \nn \\
 Q d(\bfn) &= \psi_m(\bfn)\bz_m(\bfn)
   -\bz_m(\bfn-\bfe_m)\psi_m(\bfn-\bfe_m), \nn \\
\label{BRST Q=8 old}
 Q\eta(\bfn) &=
 \frac{1}{4}\Bigl(z_m(\bfn)\bz_m(\bfn)
  -\bz_m(\bfn-\bfe_m)z_m(\bfn-\bfe_m)-d(\bfn)\Bigr),  \\
 Q\xi_{mn}(\bfn) &= \frac{1}{2}\Bigl(\bz_m(\bfn+\bfe_n)\bz_n(\bfn)
  -\bz_n(\bfn+\bfe_m)\bz_m(\bfn)\Bigr), \nn \\
 Q\chi_{lmn}(\bfn)&=0, \nn 
\end{align}
where $d(\bfn)$ is again an auxiliary field to make $Q$ nilpotent 
off-shell. 
Here we can extend (\ref{BRST Q=8 old}) by supplementing 
the fields with an additional bosonic field $f_{lmn}(\bfn)$ 
satisfying
\begin{align}
 Qf_{lmn}(\bfn)&=\chi_{lmn}(\bfn). 
\label{new field}
\end{align}
Then the action of the orbifolded matrix theory 
(\ref{orbifold action Q=8}) can be equivalently expressed 
in a $Q$-exact form:
\begin{align}
S_{\rm orb} =
\frac{1}{2g^2}{\rm Tr}\sum_{\bfn \in \Z_N^d}Q\Biggl( 
& \eta(\bfn)\Bigl({z}_m(\bfn) \bz_m(\bfn)
-\bz_m(\bfn-\bfe_m){z}_m(\bfn-\bfe_m)+d(\bfn)\Bigr) \nn \\
& - \chi_{mn}(\bfn)\Bigl(z_m(\bfn)z_n(\bfn+\bfe_n)
-z_n(\bfn)z_m(\bfn+\bfe_n)\Bigr)\nn \\
&-\frac{1}{2}f_{lmn}(\bfn)\Bigl(
\bz_l(\bfn+\bfe_m+\bfe_n)\xi_{mn}(\bfn)
-\xi_{mn}(\bfn+\bfe_l)\bz_l(\bfn)
\Bigr)\Biggr) ~. 
\label{Q-exact action Q=8}
\end{align}
Note that, although the partition function diverges by 
integration over $f_{lmn}(\bfn)$, 
it is irrelevant for the vacuum energy. 
Therefore, the 1-loop result given above is shown to be exact 
by repeating the argument in the previous subsection, 
and the vacuum energy is expected to be zero in all order
of the perturbative expansion. 

In summary, we can conclude that {\em the flat directions of the 
orbifold lattice theories constructed from the mother theory 
with four and eight supersymmetries do not receive 
any quantum correction perturbatively.}

\section{Conclusion and Discussion}

In this paper, we examined quantum corrections to 
the classical moduli space of orbifold supersymmetric lattice theories 
constructed from the $Q=4$ and $Q=8$ mother theories. 
We showed that the classical moduli space does not receive 
any quantum correction perturbatively, 
namely, the flat directions of these theories remain flat 
even if we take into account quantum effects. 
We also modified the action of the $Q=4$ 
orbifolded matrix theory 
without spoiling the $Q$-exactness and showed that the classical 
moduli space of the deformed theory does not receive 
any perturbative correction either.

We conclude this paper by making some comments on 
other orbifold lattice theories. 
Let us first consider an orbifolded matrix theory, 
\begin{align}
S_{\rm orb} = 
&\frac{1}{g^2}{\rm Tr}\sum_{\bfn}\Biggl( 
\frac{1}{4}\Bigl|z_m({\bfn})
z_n(\bfn+\bfe_m) - z_n(\bfn)z_m(\bfn+\bfe_n)\Bigr|^2 \nn \\
&+\frac{1}{8}\Bigl(z_m(\bfn)\bar{z}_m(\bfn)
 -\bar{z}_m(\bfn-\bfe_m)z_m(\bfn-\bfe_m)\Bigr)^2 \nn \\
& + \eta(\bfn)
\Bigl(\bar{z}_m(\bfn+\bfa-\bfe_m)\psi_m(\bfn+\bfa-\bfe_m)
-\psi_m(\bfn+\bfa)\bar{z}_m(\bfn+\bfa)\Bigr)\cr 
& - \frac{1}{2}\chi_{mn}(\bfn)
\Bigl(z_m(\bfn)\psi_n(\bfn+\bfe_m)
-\psi_n(\bfn)z_m(\bfn+\bfa_n) -(m \leftrightarrow n)
\Bigr)\Biggl), 
\label{general orbifold action}
\end{align}
where $\bfe_m$, $\bfa$, $\bfa_m$ and $\bfa_{12}$ are 
three-component vectors satisfying 
\begin{equation}
 \bfa+\bfa_m = \bfe_m, \quad \bfa_{12}+\bfa_m=-|\e_{mn}|\bfe_n, 
\quad \bfa+\bfa_1+\bfa_2+\bfa_{12}=0, 
\label{a-relations}
\end{equation}
$z_m(\bfn)$ and $\bz_m(\bfn)$ are the same bosonic fields as 
in (\ref{orbifold action Q=4}) but $\eta(\bfn)$, $\psi_m(\bfn)$
and $\chi_{12}(\bfn)$ are fermionic fields living on 
the links $(\bfn,\bfn+\bfa)$, $(\bfn,\bfn+\bfa_m)$ and 
$(\bfn-\bfa_{12},\bfn)$, respectively. 
In particular, we assume that any of the vectors 
$\bfa$, $\bfa_m$ and $\bfa_{12}$ 
is not zero.  
The action (\ref{general orbifold action}) 
has been first given in \cite{D'Adda:2005zk} and 
is shown to be obtained from $Q=4$ mother theory by an 
orbifold projection with no preserved supercharge 
in any usual sense \cite{Damgaard:2007eh}. 
It is easy to show that the vacuum energy 
of this theory again vanishes at the 1-loop level. 
However, in this case, there seems to be 
no guarantee that higher-loop 
contributions to the vacuum energy vanish, 
since there is no usual BRST symmetry in this theory. 
It would be interesting, however, to investigate 
quantum corrections to this theory from the view point of 
the supersymmetry algebra on lattice discussed in
\cite{D'Adda:2004jb}\cite{D'Adda:2005zk}.

Interesting orbifold lattice theories are those 
constructed from $Q=16$ mother theory \cite{Kaplan:2005ta}, 
that is, IKKT matrix theory \cite{Ishibashi:1996xs}.
The action of the corresponding orbifolded matrix theory is 
written as 
\begin{align}
S_{\rm orb} &= \frac{1}{g^2}{\rm Tr}\sum_{\bfn\in\Z_N^d}\Biggl( 
\frac{1}{4}\Bigl| z_m({\bfn})
z_n(\bfn+\bfe_m) - z_n(\bfn)z_m(\bfn+\bfe_n)\Bigr|^2  \cr
& + \frac{1}{8}\Bigl(z_m(\bfn)\bz_m(\bfn)
-\bz_m(\bfn-\bfe_m)z_m(\bfn
-\bfe_m)\Bigr)^2 \cr
& - \psi_m(\bfn)\Bigl(\bz_m(\bfn)\eta(\bfn)-\eta(\bfn+\bfe_m)
\bz_m(\bfn)\Bigr)\cr 
& +\frac{1}{2} \xi_{mn}(\bfn)\Bigl(z_m(\bfn)\psi_n(\bfn+\bfe_m)
-\psi_n(\bfn)z_m(\bfn+\bfe_n)-(m\!\leftrightarrow\!n)\Bigr) \cr
& -\frac{1}{2}\e_{mnpqr}\xi_{mn}(\bfn)\Bigl(
\bz_p(\bfn+\bfe_q+\bfe_r)\xi_{pq}(\bfn)
-\xi_{pq}(\bfn+\bfe_p)\bz_p(\bfn)
\Bigr)\Biggr)~,  
\label{orbifold action Q=16}
\end{align}
where $m,n=1,\cdots,5$, $\bfe_m$ are five-component vectors 
satisfying $\sum_{m=1}^5 \bfe_m=0$ and $d$ is the number of the 
linearly independent vectors in $\{\bfe_m\}$.  
Again the classical vacua are parametrized as (\ref{lattice vacuum})
with $m=1,\cdots 5$, and it is straightforward to show that 
the vacuum energy is zero at the 1-loop level. 
However, we cannot apply the same argument in the previous section
since the last term of the action (\ref{orbifold action Q=16}) 
is not $Q$-exact but $Q$-closed \cite{Kaplan:2005ta}. 
Thus, there is a possibility that the classical flat directions 
would be lifted up by quantum effects. 
In fact, from the viewpoint of the superstring theory, 
we can expect that non-trivial quantum corrections 
to the vacuum energy exist in this case. 
Recalling that the mother theory with sixteen supercharges is 
identical with the low energy effective theory 
on D-instantons on a ten-dimensional flat space-time,
the orbifolded matrix theory (\ref{orbifold action Q=16}) can be 
regarded as the low energy effective theory on  
D-instantons in the background of an orbifold%
\footnote{
For a related work, see \cite{Unsal:2005us}.
}. 
In this interpretation, 
the background (\ref{lattice vacuum}) can be regarded as 
the positions of D-instantons. 
The point is that this orbifold background breaks 
the supersymmetry on the ten-dimensional space-time, 
so (\ref{lattice vacuum}) or (\ref{vacuum Q=8}) gives 
a non-BPS configuration of D-branes. 
Therefore, it seems that there should be some force between 
the separated D-instantons. 
In terms of the theory on the D-instantons, 
this means the classical flat directions parametrized by 
$a_m$ are no longer flat if we take into account
quantum corrections to the orbifolded matrix model. 
It would be interesting to analyse these theories along this way
\cite{DM}.

\section*{Acknowledgments}
The author would like to thank P.~H.~Damgaard and S.~Hirano 
for useful discussions and valuable comments. 
He would also like to thank D.~B.~Kaplan, M. \"{U}nsal, 
H.~Suzuki, F.~Sugino and T.~Takimi 
for useful comments.  
This work is supported by JSPS Postdoctoral Fellowship for Research Abroad.

\bibliographystyle{JHEP}
\bibliography{refs}

\end{document}